# Single-photon interferometry and spectroscopy with two laser frequency combs

Nathalie Picqué [1,*] and Theodor W. Hänsch [1,2]

1. Max-Planck Institute of Quantum Optics
Hans-Kopfermann-Str. 1, 85748 Garching, Germany
2. Ludwig-Maximilian University of Munich, Faculty of Physics,
Schellingstr. 4, 80799 Munich, Germany
* nathalie.picque@mpq.mpg.de

**Abstract**
We demonstrate single-photon time-domain interference in a new realm. We observe interferences in the photon counting statistics with two separate mode-locked femtosecond lasers of slightly different repetition frequencies, each emitting a comb of evenly spaced spectral lines over a wide spectral span. We exploit the interference pattern for spectroscopic diagnostics over a broad spectral range. An experimental proof-of-concept shows that the emerging technique of high-resolution dual-comb Fourier transform spectroscopy can be performed at light powers that are a billion-fold weaker than those commonly employed. Our experiments challenge the intuitive concept that a photon exists before detection and they open the prospect of precise spectroscopy over broad spectral bandwidth in light-starved conditions.

We demonstrate broadband dual-comb spectroscopy with two femtosecond laser frequency combs even when the lasers are so strongly attenuated that only single photons are detected. The detection rates are so low that it is extremely unlikely that two photons, one from each source, are present in the detection path at the same time.
Dual-comb spectroscopy with two mode-locked lasers of slightly different repetition frequencies is emerging as a powerful tool for broadband spectroscopy with high precision [1]. A number of demonstrations harnessing a variety of laser sources [2-8] and sampling techniques [9-12] show a convincing potential. The principle is easily described in terms of classical electromagnetic waves. Each laser emits a large number of precisely equally spaced spectral comb lines [13]. Pairs of comb lines, one from each laser source, interfere at a fast photodetector, so that a comb of radio-frequency beat notes appears in the detector signal, where it is accessible to digital signal processing. Any optical spectral structure introduced by a sample is transformed to a corresponding structure in the radio-frequency spectrum. Optical resonance frequencies in a sample are effectively slowed down by a large factor equal to the repetition frequency divided by the difference in repetition frequencies.
Here we demonstrate dual-comb spectroscopy at such low light levels that only one single photon is detected on average during the time of a thousand laser pulses. The statistics of the detected single photons carries the information about the two femtosecond lasers with their hundreds of thousands of comb lines and about the sample with its possibly highly complex optical spectrum.
For the proof-of-principle experiments reported here (Fig.1, Appendix), two essentially free-running mode-locked erbium-doped fiber femtosecond lasers with second harmonic generation to 780 nm are used. Their repetition frequency is about 100 MHz, corresponding to a period of 10 ns, and the difference in repetition frequencies is maintained at a chosen value between 10 Hz and 300 Hz using a slow feedback loop. The beams of the two lasers are combined on a pellicle beamsplitter. One output of the interferometer is strongly attenuated to the picowatt or femtowatt range as detailed below. A solid étalon with silvered end faces and a free spectral range of 500 GHz is used as a sample. The photons are counted on a single-photon detector and the counts are acquired as a function of time by a multiscaler. To obtain a dual-comb single-photon interferogram, which approaches its classical counterpart, many photon clicks have to be accumulated. Here we use a simple yet limited approach which consists in triggering the data acquisition using the central fringe, detected with a standard fast silicon photodiode, at the second output of the interferometer and in accumulating many scans that add up the clicks for each time delay per increments of 800 ps.

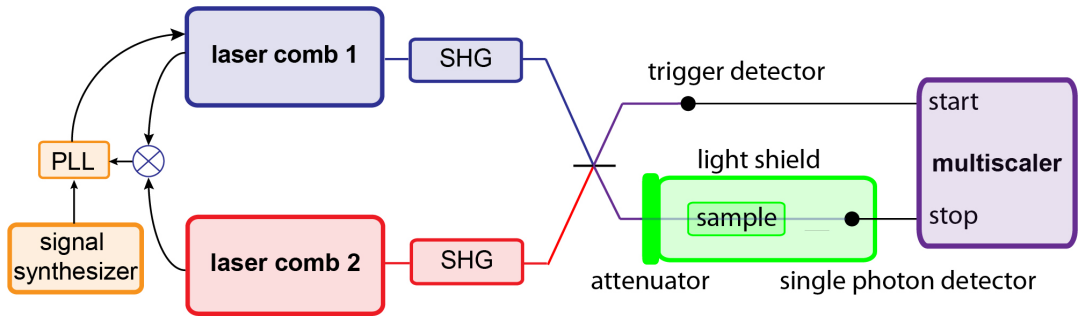

**Figure 1. Sketch of the experimental set-up for single-photon dual-comb spectroscopy.**
Two femtosecond laser oscillators, labeled laser comb 1 and laser comb 2, have a stabilized difference $\Delta f_{rep}$ in repetition frequencies. Their frequency-doubled beams with a spectrum centered at 780 nm (384 THz), are combined on a beam-splitter. The fringes at one output of the interferometer provides a trigger signal. The second output of the interferometer is strongly attenuated to a flux of photons lower than $3 \times 10^6$ photon s$^{-1}$, implying that the mean interval between photons is always more than one-hundred-fold longer than the transit time of a photon in the interferometer. The photon clicks are counted by a single-photon detector and accumulated as a function of arrival time after the trigger by a multi-scaler.

The interferogram shown in Fig. 2a (see also Appendix) has been recorded at an average rate of one photon every 368 ns, corresponding to an average power at the detector of $1.1 \times 10^{-12}$ W. A total of $1.4 \times 10^6$ clicks are detected in $1.2 \times 10^5$ scans, of 4.25 µs each, in a total experimental time of 10 minutes. The interferogram shows a non-interferential background, which is usually electronically filtered out in dual-comb spectroscopy and that we numerically filter in Fig. 2b. Superimposed on the background signal, five bursts with good modulation contrast are discernible. They correspond to the interference pattern at the output of the Fabry-Pérot étalon. The Fourier transform of the interferogram reveals the spectrum of the étalon (Fig.2c), downconverted in the radio-frequency region. In the optical region, the spectrum spans about 7 THz.

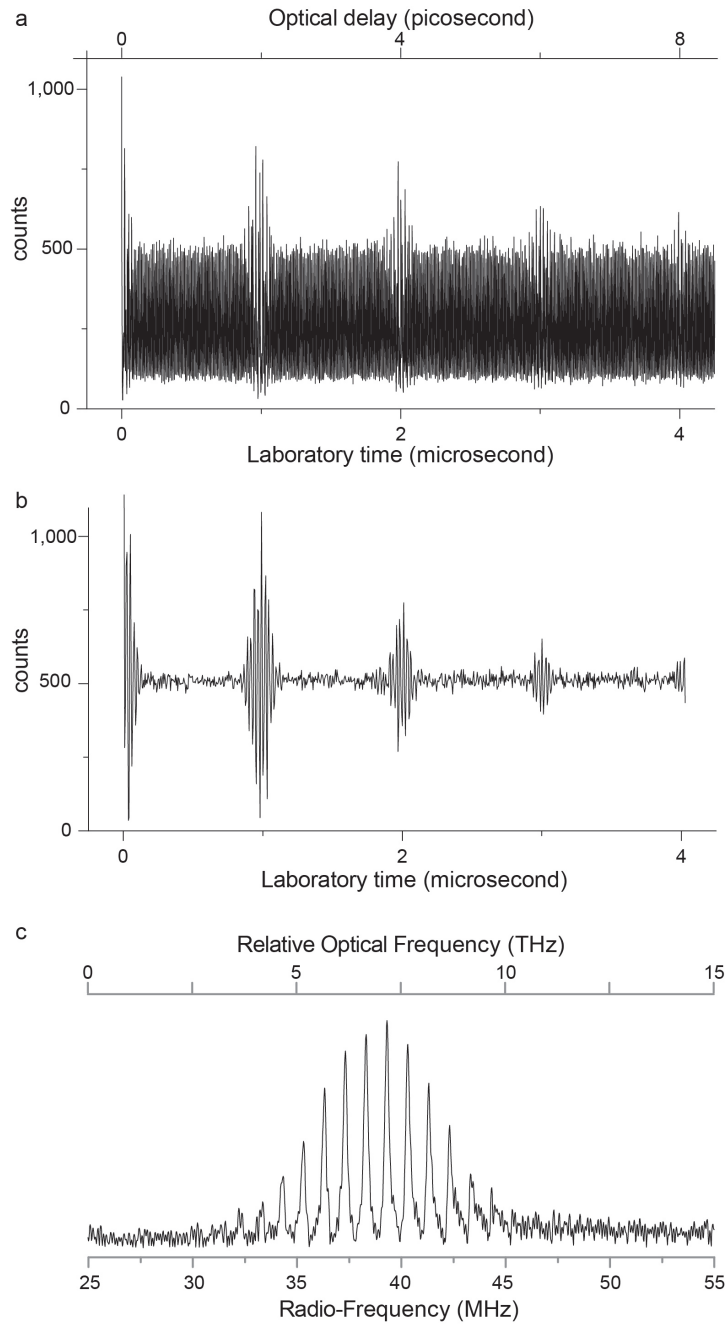

**Figure 2. Experimental dual-comb interferogram and spectrum at a detected photon rate 2.7 $10^6$ photon s$^{-1}$.**
a. The raw interferogram sampled at a rate of 0.8 ns (lower x-scale: laboratory time) comprises a background part, unmodulated by the interference. The periodic interference signatures of a Fabry-Pérot étalon appear with a high contrast even at an optical delay of 8 ps (upper x-scale: optical delays).
b. The interferogram of a. is numerically filtered to highlight better the single-photon interference signal.
c. The Fourier transform of a single-photon interferogram (including a.) reveals the evenly-spaced transmission lines of the étalon, over a broad spectral span of several THz in the optical domain.

Lower light levels are reachable. An interferogram of the same Fabry-Pérot étalon is recorded (Fig.3a) at an average power as low as $20 \cdot 10^{-15}$ W. This corresponds to an average rate of one photon every 21 µs, or one photon during the time of two thousands pulse pairs, or 0.14 photon per interferometric scan of 3.0 µs. The Fourier transform of the interferogram properly reproduces the spectrum of the étalon (Fig.3b). In experiments of dual-comb spectroscopy, the average power at the detector is usually on the order of a few tens of microwatt. We thus achieve a billion-fold decrease of the light level using single-photon counting technology.

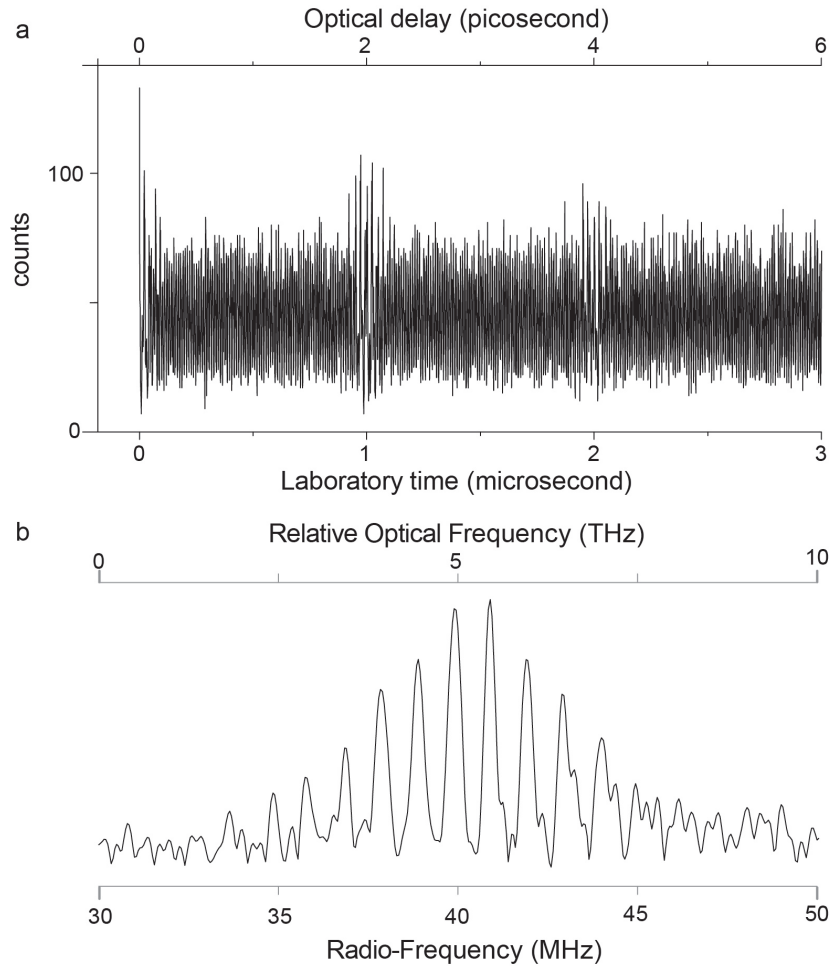

**Figure 3. Experimental dual-comb interferogram and spectrum under an average power of 20 femtowatt at the detector.**
a. Interferogram comprising only 168703 detected clicks
b. Spectrum, Fourier transform of the interferogram, retrieved at a power level one billion-fold lower than usual dual-comb spectra.

Single photon interference fringes have long been observed in Young's double slit experiment or with a Michelson/Mach-Zehnder interferometer [14] and have continuously attracted interest, as exemplified by a recent demonstration with a frequency comb [15]. In such a situation, it is possible to imagine that a photon from a defined source can follow two different quantum paths, which interfere constructively or destructively at the detector. Richard Feynman famously called such interference "the only mystery" in quantum physics [16]. For our dual-comb interferometry with two separate laser sources, we have to abandon the concept that a photon exists before detection. The photon is defined by the detection event. To predict the probability of such an event, one has to sum the probability amplitudes from any quantum paths that can contribute to this detector click.

The possibility of observing interference effects by detecting single photons from two mutually coherent continuous-wave single-mode laser sources has first been demonstrated in 1967 [17]. The experiments presented here extend this phenomenon to the much more complex scenario of two mode-locked femtosecond lasers and broadband spectroscopy.

One might argue that no dual-comb interferences can be expected with single photons from two femtosecond lasers, since the arrival time of a photon makes it possible to determine which of the two lasers has emitted the photon. To counter this argument, it is useful to look at dual comb spectroscopy in the time domain. A single short pulse that interacts with narrow spectral resonances of a sample acquires a temporal tail, such as the free induction decay field due to molecular resonances. This tail carries most of the spectral information and is read out interferometrically in Fourier transform spectroscopy or dual-comb spectroscopy. As long as the temporal separation of two pulses does not exceed the coherence time of the sample, a photon click cannot be uniquely assigned to one particular laser, and interference between two photon paths originating from different sources becomes possible.

The proof-of-concept described here, although fully demonstrative of single photon interference between two femtosecond frequency comb generators, is technically limited by our elementary experimental implementation. For interferometry, the mutual coherence of the two laser sources has to be maintained during the entire recording time. Sophisticated electronic stabilization techniques have been developed to maintain mutual coherence of two femtosecond lasers over periods of seconds. Recently, a feed-forward dual-comb spectrometer with a experimental mutual coherence time exceeding thirty minutes has been demonstrated [7]. This latter development has actually motivated the present proof-of-principle as it enables the straightforward use of single-photon counting instrumentation with dual-comb spectroscopy by permitting direct accumulation of photon clicks over extended time period. Now that the feasibility of single-photon dual-comb interferometry is established, forthcoming investigations will harness feed-forward stabilization to enable scientific applications through long term averaging. With further system development, a new spectroscopic tool may become available for studying fundamental optical processes at the level of single photons and single emitters.

Aside from its conceptual interest, dual-comb spectroscopy at the single photon level may also open new opportunities for experiments in spectral regions, where only low light levels are available, e.g. after nonlinear frequency conversion, in fluorescence detection of few atoms/molecules, or in situations where strong attenuation cannot be avoided, such as in highly scattering media or in remote sensing. For instance, Fourier transform excitation spectroscopy of single molecular emitters shows an intriguing potential for revealing the interaction between the molecules and their local nano-scale environment [18-20]. Linear and nonlinear [21] single-molecule spectroscopy under dual-comb excitation [22] could provide the advantages of a multimodal spectrometer without moving parts of an excellent accuracy in measuring the shapes and positions [1] of the spectral lines of single emitters at high resolution. Another prospect addresses the different scientific challenge of tests of quantum molecular theory [23]. For precision spectroscopy of simple molecules, dual-comb spectroscopy now holds out the promise of combined wide spectral bandwidth and sub-Doppler resolution [22] in a controlled environment of rarefied or cold molecules or of low photon fluxes.

The experiments that we present towards dual-comb spectroscopy rely on first order interference, which is robust against attenuation. More sophisticated but also more delicate approaches to spectroscopy may become possible by exploiting higher order correlations and the photon entanglement that can be created with quantum optical frequency combs [24-27].

## Acknowledgements.

We are grateful to Anton Scheich for the design and implementation of locking electronics. Work performed within the frame of the Munich Center for Quantum Science and Technology funded by the Deutsche Forschungsgemeinschaft (DFG, German Research Foundation) under Germany's Excellence Strategy – EXC-2111 – 390814868. Support by the Carl-Friedrich von Siemens Foundation is acknowledged.

## Appendix.

### Detailed experimental set-up

Two amplified erbium-doped fiber mode-locked lasers (Fig.1) emit pulses at a repetition frequency of about $f_{rep}$ = 100 MHz and an average power of 220 mW. Their spectrum is centered at 1560 nm (192 THz). They are frequency doubled to produce a spectrum centered at 780 nm (384 THz), with an average power of 50 mW. In the experiments reported here, we mainly rely on the passive mechanical stability of the lasers. The temperatures of the two lasers are controlled with Peltier elements and this enables coarse setting of the repetition frequency. Only the difference of the repetition frequencies is actively stabilized to a frequency-synthesized signal generator, at a value $\Delta f_{rep}$ chosen between 10 Hz and 300 Hz, with the help of a slow digital phase-locked loop. The error signal from the phase-locked loop is used to adjust the voltage for a fiber-stretching piezo-electric element inside one of the lasers, so that the difference in pulse repetition frequencies can be maintained at the chosen value. The repetition frequency of the other laser and the carrier-envelope frequencies of the two lasers are left unstabilized. Regardless of the fluctuations, not stabilizing the difference in carrier-envelope frequencies to a well-chosen value fundamentally limits the capabilities of time-domain accumulation of the interferometric signal, as an arbitrary value of this difference implies that the time-domain interferometric signal does not periodically recur.

The beams of the two frequency-doubled femtosecond lasers are superimposed with a pellicle beam-splitter. One output of the beam-splitter is sent, after suitable attenuation, to a fast silicon photodiode. When the pulses of the two femtosecond lasers overlap in time, a strong interference signal is produced. Its central fringe is used to generate the trigger signal, which starts the photon counting by a multi-scaler. The second output beam is used for single-photon dual-comb spectroscopy. To avoid aliasing in the dual comb spectra at the chosen repetition rate difference of 200 Hz, the frequency-doubled laser beams are first sent through a bandpass filter, which narrows the spectral span to about 7 THz and stretches the pulses to about 300 fs.

After attenuation to a flux between $10^4$ to $3 \times 10^6$ photon counts per second, it passes through a sample. In this simple proof-of-concept, a solid étalon with silvered end faces and a free spectral range of 500 GHz is used as sample. The beam is then directed to a fiber-coupled single-photon detector of low dark count rate. A suitable light shield protects the detector input from ambient stray light. After the multi-scaler receives a trigger "start" signal, any single photon signals received by the "stop" input are counted according to their arrival time in time-bins with increments of 800 ps.

**Detailed recording conditions**

In the recording leading to the data shown in Figure 2a, a total of 1406790 photons have been counted over 121894 sweeps of 5312 time bins of 0.8 ns each. The duration of one sweep is 4.25 µs and the total measurement time is 0.52 seconds. The cycling time of a sweep is $1/\Delta f_{rep} = 1/(200\ Hz) = 5$ ms, therefore the entire time for the experiment is 10 minutes. Unfortunately, insufficient stability of the difference of laser repetition rates is so far preventing the averaging of such interferograms over times longer than a few minutes, except for the central fringe pattern and the first three or four subsequent interference bursts produced by the étalon. Nevertheless, the clear interference contrast unambiguously demonstrates that dual-comb interferograms can be accumulated by photon counting even at very low light levels. On average, 11.5 clicks per sweep of 4.25 µs are measured. The detected photon rate is therefore $2.72\ 10^6$ photon per second. In average, 1 photon every 368 ns is detected, whereas the transit time of a photon from the sources to the detector is about 3 ns. The photons have a wavelength of 780 nm. The average power corresponding to the detected interferogram is 690 femtowatt. The efficiency of the detector is 60% and the detector dead time after a count is 43 ns, which bring the power falling onto the detector to $1.15\ 10^{-12}$ W. The interferogram extends to 10000 time bins and it is completed with zeros up to sixty thousand samples in order to interpolate the spectrum, as is common practice in Fourier transform spectroscopy. A magnitude Fourier transform is computed. The resulting radio-frequency spectrum of the étalon shown in Fig. 2c has a resolution of 125 kHz. In dual-comb spectroscopy, the optical frequencies are down-converted to the radio-frequency domain by a factor $\Delta f_{rep}/f_{rep}$. A resolution of 125 kHz in the radio-frequency domain corresponds, in the optical domain, to a resolution of 125 kHz x 100 MHz / 200 Hz, that is 62.5 GHz. The central optical frequency in the spectrum is about 384 THz. In this experiment, absolute calibration of the frequency scale is not achieved, as the carrier-envelope frequencies of the lasers are not measured. State-of-the art dual-comb set-ups, though, enable absolute frequency calibration [6, 7, 22] and we will implement this functionality together with long coherence times [7] in future work. For now, we only display the relative optical scale in the upper x-axis of Fig. 2c.

At low light level, the cumulated photon count over 3750 time bins of 0.8 ns each is shown in Fig. 3a. A total of 168703 photons have been counted over 1190033 sweeps, each lasting 3.0 µs. On average 0.14 clicks are captured per sweep of 3.0 µs. The total acquisition time is 3.57 seconds while, owing to the interferogram cycling time, the experiment lasts 1 hour and 39 minutes. The detected photon rate is 47254 photon s$^{-1}$. In other words, on average, one photon is detected every 21 µs. The average power corresponding to the detected interferogram is $1.2\ 10^{-14}$ W = 12 femtowatt. Accounting for the detector efficiency, the power falling onto the detector is $2\ 10^{-14}$ W. An interferogram extended to 4376 time bins is completed with five-fold more zeros than actual samples for spectral interpolation and it is Fourier transform to reveal the spectrum shown in Fig. 3b with a radio-frequency resolution of 285 kHz, corresponding to 143 GHz in the optical domain. The spectrum in Fig. 3b appears centered at a different radio-frequency than that of Fig. 2c because of drifts, and the absence of control, of the carrier-envelope offset frequencies of the lasers. It is however centered at the same optical frequency as that of Fig. 2c.

# References

[1] N. Picqué, T.W. Hänsch, Frequency comb spectroscopy, *Nature Photonics* **13**, 146-157 (2019).
[2] G. Villares, A. Hugi, S. Blaser, J. Faist, Dual-comb spectroscopy based on quantum-cascade-laser frequency combs, *Nature Communications* **5**, 5192 (2014).
[3] M.-G. Suh, Q.-F. Yang, K.Y. Yang, X. Yi, K.J. Vahala, Microresonator soliton dual-comb spectroscopy, *Science* **354**, 600-603 (2016).
[4] G. Millot, S. Pitois, M. Yan, T. Hovhannisyan, A. Bendahmane, T.W. Hänsch, N. Picqué, Frequency-agile dual-comb spectroscopy, *Nature Photonics* **10**, 27-30 (2016).
[5] S.M. Link, D.J.H.C. Maas, D. Waldburger, U. Keller, Dual-comb spectroscopy of water vapor with a free-running semiconductor disk laser, *Science* **356**, 1164-1168 (2017).
[6] G. Ycas, F.R. Giorgetta, E. Baumann, I. Coddington, D. Herman, S.A. Diddams, N.R. Newbury, High-coherence mid-infrared dual-comb spectroscopy spanning 2.6 to 5.2 mu m, *Nature Photonics* **12**, 202-208 (2018).
[7] Z. Chen, M. Yan, T.W. Hänsch, N. Picqué, A phase-stable dual-comb interferometer, *Nature Communications* **9**, 3035 (2018).
[8] A. Dutt, C. Joshi, X.C. Ji, J. Cardenas, Y. Okawachi, K. Luke, A.L. Gaeta, M. Lipson, On-chip dual-comb source for spectroscopy, *Science Advances* **4**, e1701858 (2018).
[9] B. Bernhardt, A. Ozawa, P. Jacquet, M. Jacquey, Y. Kobayashi, T. Udem, R. Holzwarth, G. Guelachvili, T.W. Hänsch, N. Picqué, Cavity-enhanced dual-comb spectroscopy, *Nature Photonics* **4**, 55-57 (2010).
[10] T. Ideguchi, S. Holzner, B. Bernhardt, G. Guelachvili, N. Picqué, T.W. Hänsch, Coherent Raman spectro-imaging with laser frequency combs, *Nature* **502**, 355-358 (2013).
[11] B. Lomsadze, S.T. Cundiff, Frequency combs enable rapid and high-resolution multidimensional coherent spectroscopy, *Science* **357**, 1389-1391 (2017).
[12] T. Minamikawa, Y.-D. Hsieh, K. Shibuya, E. Hase, Y. Kaneoka, S. Okubo, H. Inaba, Y. Mizutani, H. Yamamoto, T. Iwata, T. Yasui, Dual-comb spectroscopic ellipsometry, *Nature Communications* **8**, 610 (2017).
[13] T.W. Hänsch, Nobel Lecture: Passion for precision, *Reviews of Modern Physics* **78**, 1297-1309 (2006).
[14] P. Grangier, G. Roger, A. Aspect, Experimental Evidence for a Photon Anticorrelation Effect on a Beam Splitter: A New Light on Single-Photon Interferences, *Europhysics Letters (EPL)* **1**, 173-179 (1986).
[15] S.K. Lee, N.S. Han, T.H. Yoon, M. Cho, Frequency comb single-photon interferometry, *Communications Physics* **1**, 51 (2018).

[16] R. Feynman, R. Leighton, M. Sands, The Feynman Lectures on Physics, Volume III: Quantum Mechanics, New Millennium ed., Basic Books, New York, NY, 2011.

[17] R.L. Pfleegor, L. Mandel, Interference of Independent Photon Beams, *Physical Review* **159**, 1084-1088 (1967).

[18] R. Korlacki, M. Steiner, H. Qian, A. Hartschuh, A.J. Meixner, Optical Fourier Transform Spectroscopy of Single-Walled Carbon Nanotubes and Single Molecules, *ChemPhysChem* **8**, 1049-1055 (2007).

[19] L. Piatkowski, E. Gellings, N.F. van Hulst, Broadband single-molecule excitation spectroscopy, *Nature Communications* **7**, 10411 (2016).

[20] E. Thyrhaug, S. Krause, A. Perri, G. Cerullo, D. Polli, T. Vosch, J. Hauer, Single-molecule excitation–emission spectroscopy, *Proceedings of the National Academy of Sciences* **116**, 4064 (2019).

[21] A. Maser, B. Gmeiner, T. Utikal, S. Götzinger, V. Sandoghdar, Few-photon coherent nonlinear optics with a single molecule, *Nature Photonics* **10**, 450 (2016).

[22] S.A. Meek, A. Hipke, G. Guelachvili, T.W. Hänsch, N. Picqué, Doppler-free Fourier transform spectroscopy, *Optics Letters* **43**, 162-165 (2018).

[23] N. Hölsch, M. Beyer, E.J. Salumbides, K.S.E. Eikema, W. Ubachs, C. Jungen, F. Merkt, Benchmarking Theory with an Improved Measurement of the Ionization and Dissociation Energies of H2, *Physical Review Letters* **122**, 103002 (2019).

[24] J. Roslund, R.M. de Araújo, S. Jiang, C. Fabre, N. Treps, Wavelength-multiplexed quantum networks with ultrafast frequency combs, *Nature Photonics* **8**, 109 (2013).

[25] M. Chen, N.C. Menicucci, O. Pfister, Experimental Realization of Multipartite Entanglement of 60 Modes of a Quantum Optical Frequency Comb, *Physical Review Letters* **112**, 120505 (2014).

[26] Z. Xie, T. Zhong, S. Shrestha, X. Xu, J. Liang, Y.-X. Gong, J.C. Bienfang, A. Restelli, J.H. Shapiro, F.N.C. Wong, C. Wei Wong, Harnessing high-dimensional hyperentanglement through a biphoton frequency comb, *Nature Photonics* **9**, 536 (2015).

[27] C. Reimer, M. Kues, P. Roztocki, B. Wetzel, F. Grazioso, B.E. Little, S.T. Chu, T. Johnston, Y. Bromberg, L. Caspani, D.J. Moss, R. Morandotti, Generation of multiphoton entangled quantum states by means of integrated frequency combs, *Science* **351**, 1176-1180 (2016).